\documentclass[a4paper]{spie}  

\usepackage[english]{babel}

\usepackage{amsmath,amsfonts,amssymb}
\usepackage{graphicx}
\usepackage[colorlinks=true, allcolors=blue]{hyperref}
\usepackage{subcaption}
\usepackage{geometry}
\title{The internal alignment and validation of a powered ADC for SOXS}
\author[a]{F. Battaini}
\author[a]{K. Radhakrishnan}
\author[a]{R. Claudi}
\author[b]{M. Munari}
\author[b]{R.Z. Sánchez}
\author[c]{M. Aliverti}
\author[d]{M. Colapietro}
\author[a]{D. Ricci}
\author[a]{L. Lessio}
\author[a]{M. Dima}
\author[e]{F. Biondi}
\author[c]{S. Campana}
\author[d]{P. Schipani}
\author[a]{A. Baruffolo}
\author[f]{S. Ben-Ami}
\author[d]{G. Capasso}
\author[g]{R. Cosentino}
\author[h]{F. D’Alessio}
\author[c]{P. D’Avanzo}
\author[f]{O. Hershkod}
\author[i]{H. Kuncarayakti}
\author[c]{M. Landoni}
\author[l]{G. Pignata}
\author[m]{A. Rubin}
\author[n]{S. Scuderi}
\author[h]{F. Vitali}
\author[o]{D. Young}
\author[p]{J. Achrén}
\author[q]{J. A. Araiza-Duran}
\author[r]{I. Arcavi}
\author[q]{A. Brucalassi}
\author[f]{R. Bruch}
\author[a]{E. Cappellaro}
\author[d]{M. Della Valle}
\author[a]{M. De Pascale}
\author[b]{R. Di Benedetto}
\author[d]{S. D’Orsi}
\author[f]{A. Gal-Yam}
\author[c]{M. Genoni}
\author[g]{M. Hernandez}
\author[r]{J. Kotilainen}
\author[s]{G. Li Causi}
\author[c]{L. Marty}
\author[i]{S. Mattila}
\author[f]{M. Rappaportd}
\author[c]{M. Riva}
\author[a]{B. Salasnich}
\author[o]{S. Smartt}
\author[t]{M. Stritzingerv}
\author[g]{H. Venturae}

\affil[a]{INAF - Osservatorio Astronomico di Padova, Padova, Italy}
\affil[b]{INAF - Osservatorio Astrofisico di Catania, Catania, Italy}
\affil[c]{INAF - Osservatorio Astronomico di Brera, Merate, Italy}
\affil[d]{INAF - Osservatorio Astronomico di Capodimonte, Napoli, Italy}
\affil[e]{Max-Planck-Institut für Extraterrestrische Physik, Garching, Germany}
\affil[f]{Weizmann Institute of Science, Rehovot, Israel}
\affil[g]{INAF - Fundación Galileo Galilei, Breña Baja, Spain}
\affil[h]{INAF - Osservatorio Astronomico di Roma, Roma, Italy}
\affil[i]{Tuorla Observatory, Department of Physics and Astronomy, University of Turku, Finland}
\affil[l]{Universidad Andres Bello, Santiago, Chile}
\affil[m]{European Southern Observatory, Garching, Germany}
\affil[n]{INAF - Istituto di Astrofisica Spaziale e Fisica Cosmica, Milano, Italy}
\affil[o]{Queen's University Belfast, School of Mathematics and Physics, Belfast, UK}
\affil[p]{Incident Angle Oy, Turku, Finland}
\affil[q]{INAF-Osservatorio Astrofisico di Arcetri, Firenze, Italy}
\affil[r]{Tel Aviv University, Tel Aviv, Israel}
\affil[s]{INAF - Istituto di Astrofisica e Planetologia Spaziali, Roma, Italy}
\affil[t]{Aarhus University, Aarhus, Denmark}

\authorinfo{Send correspondence to federico.battaini@inaf.it}

\begin{document} 
\maketitle

\begin{abstract}
The Son Of X-Shooter (SOXS) is a two-channel spectrograph along with imaging capabilities, characterized by a wide spectral coverage (350nm to 2000nm), designed for the NTT telescope at the La Silla Observatory. Its main scientific goal is the spectroscopic follow-up of transients and variable objects. The UV-VIS arm, of the Common Path sub-system, is characterized by the presence of a powered Atmospheric Dispersion Corrector composed (ADC) by two counter-rotating quadruplets, two prisms, and two lenses each. The presence of powered optics in both the optical groups represents an additional challenge in the alignment procedures. We present the characteristics of the ADC, the analysis after receiving the optics from the manufacturer, the emerging issues, the alignment strategies we followed, and the final results of the ADC in dispersion and optical quality.
\end{abstract}


\section{INTRODUCTION}
\label{sec:intro}  

\begin{figure}
    \centering
    \includegraphics[width=0.9\textwidth]{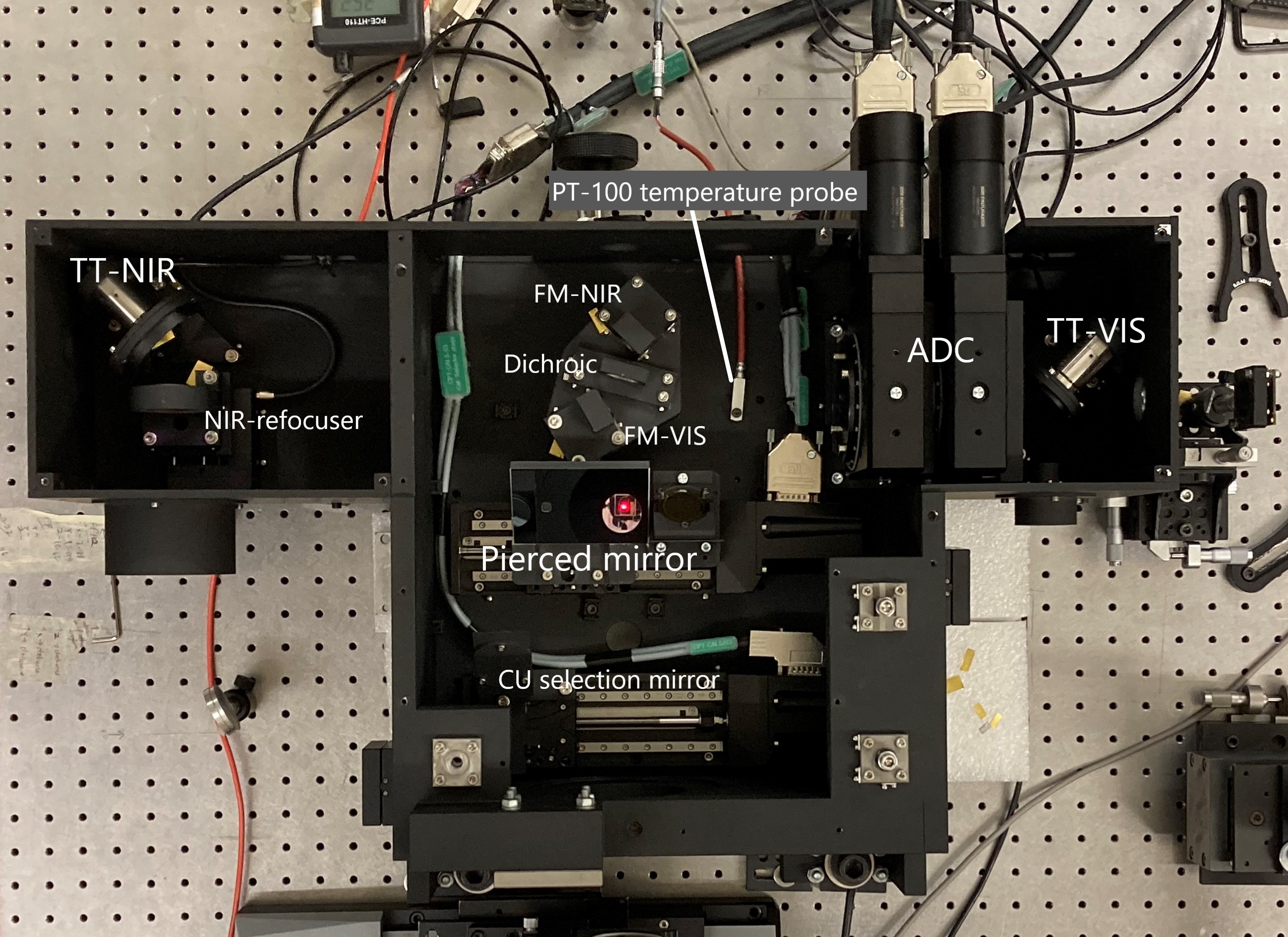}
    \caption{The Common Path is the first subsystem of SOXS, it recives the light from the telescope (F/11) and send it to the UV-VIS and NIR spectrograph or to the Acquisition Camera}
    \label{fig:commonpath}
\end{figure}

SOXS (Son Of X-Shooter) is a two-channel spectrograph, spanning UV-VIS (350-850nm) and NIR (800-2000nm) wavelength regimes\cite{Schipani16}. The Common Path sub-system receives the F/11 telescope beam and splits the UV-VIS and NIR wavelengths and provide F/6.5 beams to the respective spectrographs\cite{Claudi18}. In figure \ref{fig:commonpath} is possible to see the Common Path fully populated. 

\begin{figure}
    \centering
    \includegraphics[width=0.6\textwidth]{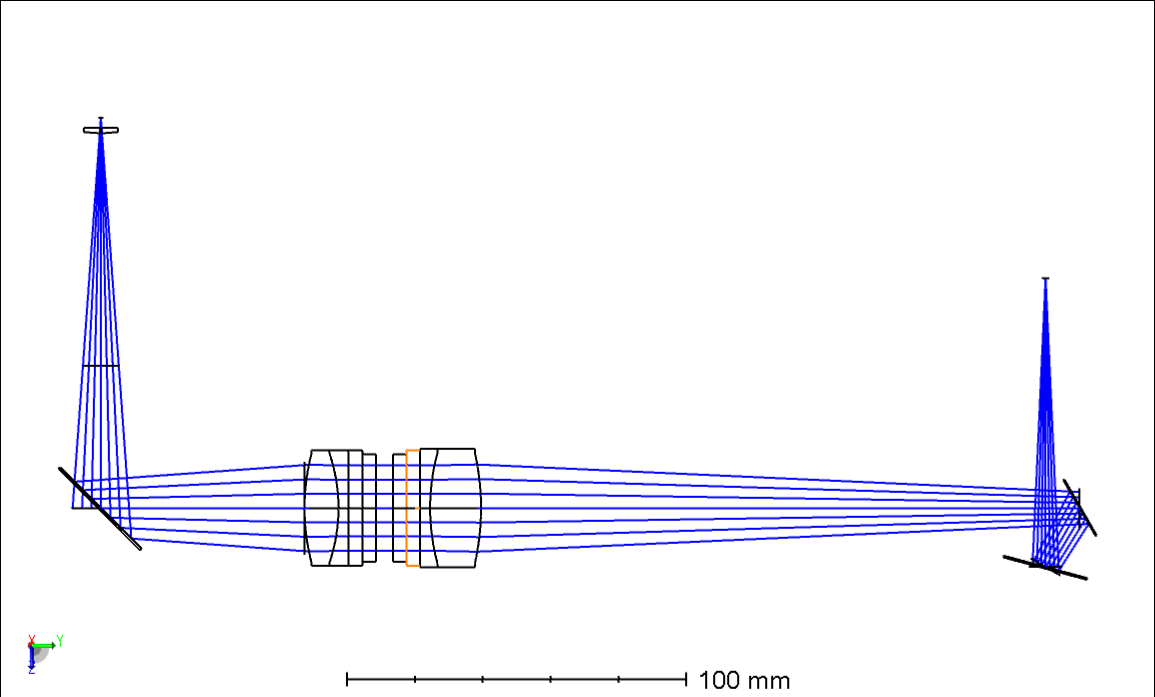}
    \caption{Optical path of the UVVIS arm of the Common Path of SOXS, view from the top}
    \label{fig:CPvis}
\end{figure}

\begin{figure}
    \centering
    \includegraphics[width=0.6\textwidth]{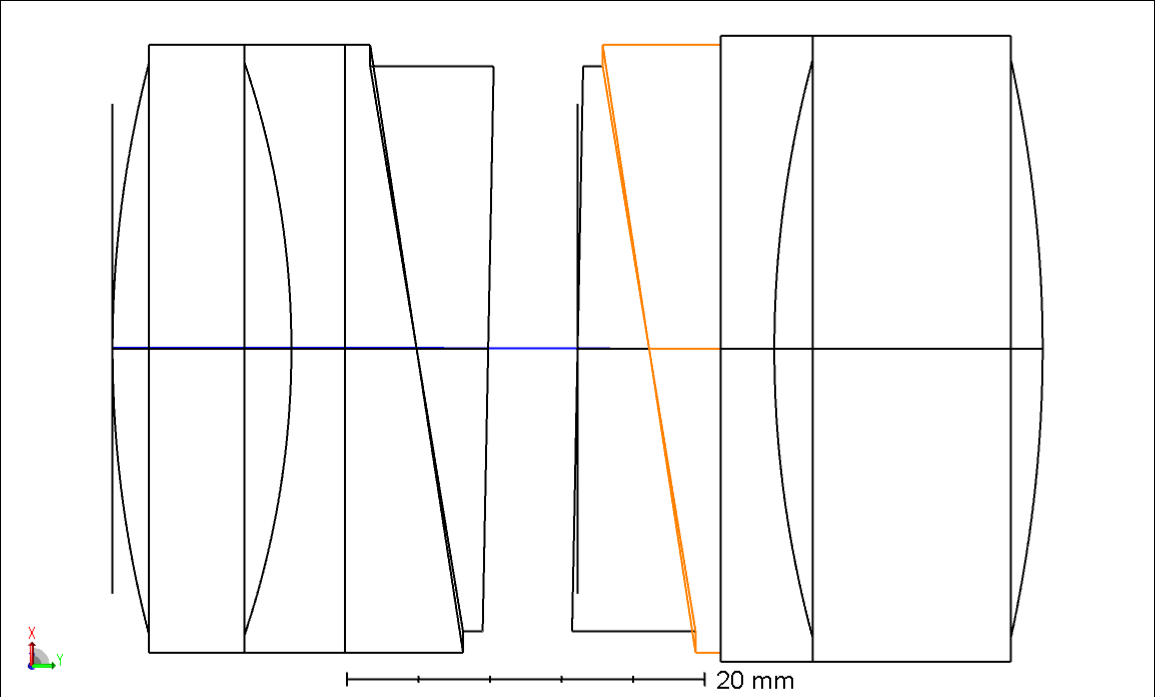}
    \caption{Scheme of the SOXS's ADC formed by two quadruplets. On the right the ADC1.}
    \label{fig:ADCscheme}
\end{figure}

The UV-VIS arm (see Fig. \ref{fig:CPvis}) is characterized by the presence of a powered Atmospheric Dispersion Corrector (ADC) that includes both the dispersion elements and optics to reduce the f-number of the beam. The SOXS ADC composed of two counter-rotating quadruplets: each of them has two prisms and two lenses as show in Fig. \ref{fig:ADCscheme}\cite{Sanchez18}.
ArcherOptx manufactured and delivered the two quadruplets inserted in the respective barrels and motors. 
In the AIV process\cite{Biondi20},our analysis suggested re-alignment of the ADC unit to match the required optical performances. The presence of powered optics in both the ADC’s optical groups translates into need for tight tolerances. In order to proceed with the new alignment and to face some possible issues with glue used by the manufacturer we unglued the two quadruplets from their barrels, we performed a temporary alignment and finally we glued it again. 

\section{ALIGNMENT EVALUATION ON RECEIVING}

\begin{figure}
    \centering
    \begin{subfigure}[b]{0.35\textwidth}
        \centering
        \includegraphics[width=\textwidth]{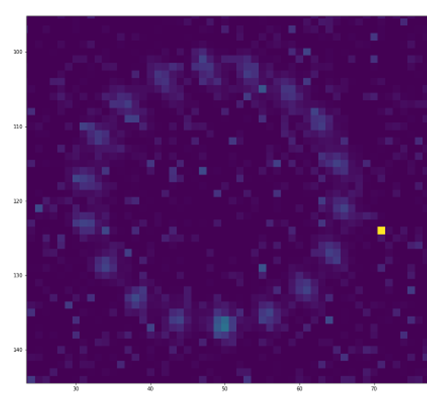}
        \caption{ADC2 $150\mu m$ diameter}
    \end{subfigure}
    \begin{subfigure}[b]{0.35\textwidth}
        \centering
        \includegraphics[width=\textwidth]{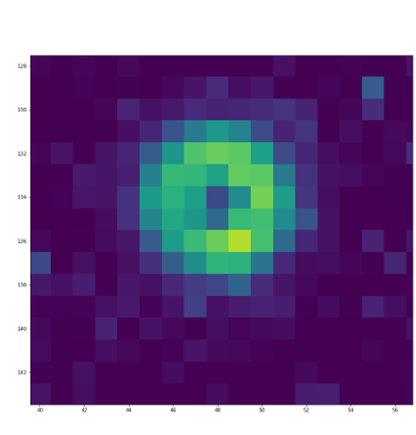}
        \caption{ADC1 $30\mu m$ diameter}
    \end{subfigure}
\caption{The effect of rotating ADC1 or ADC2 at the moment of receiving}
\label{fig:rotationsatrecieving}
\end{figure}

\begin{figure}
    \centering
    \includegraphics{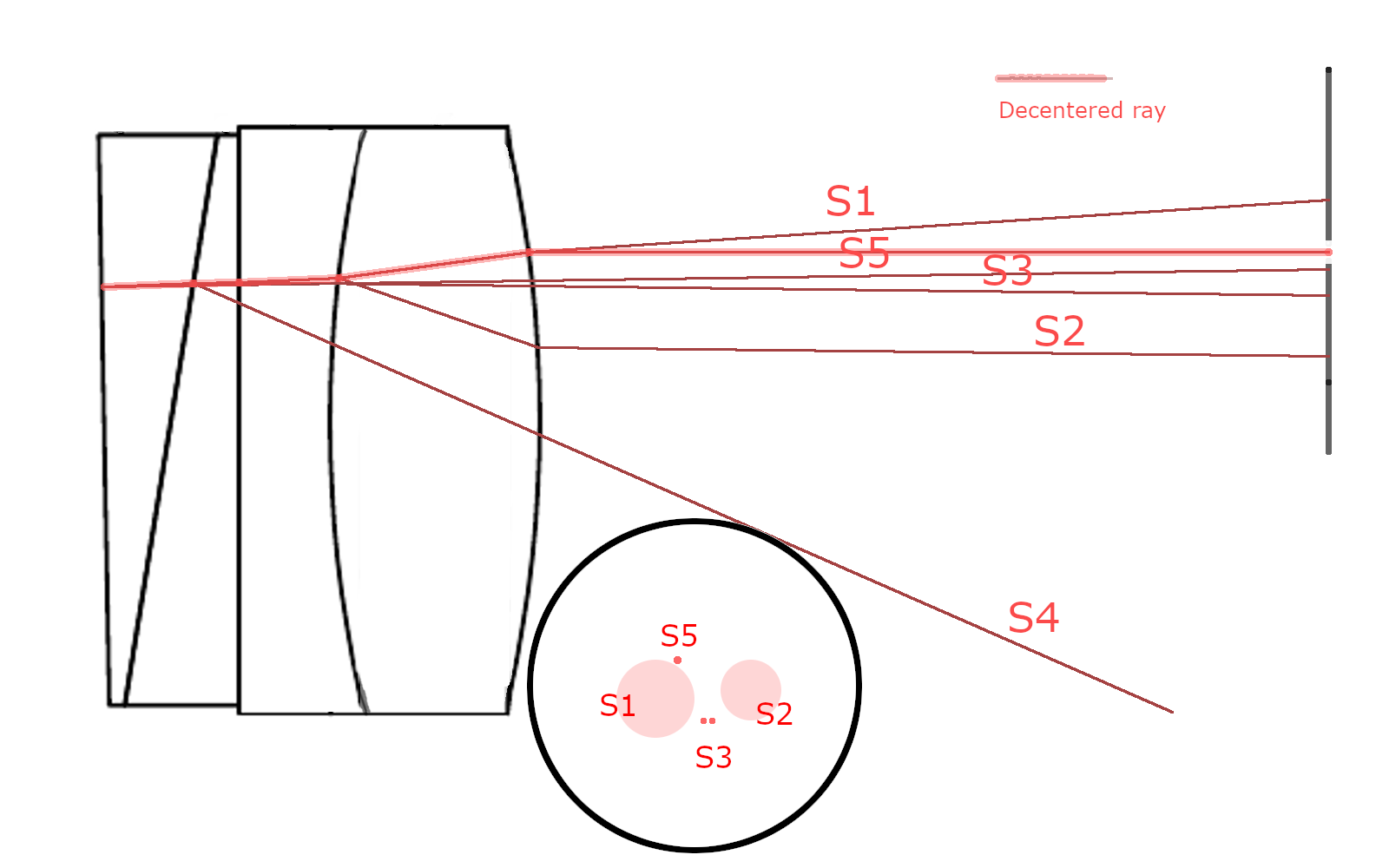}
    \caption{The reflections from the surfaces of one of the ADC while the beam is decentered, it was possible to address each reflection and to use them to check the alignment, rotating the system}
    \label{fig:reflections}
\end{figure}

We performed a series of tests to verify and validate the ADC mounting and alignment. Note that we name the first, in the optical path, quadruplet unit and the motor as ADC1, whose first lens is a CaF2 one, and the other as ADC2.\\
As the first step, we placed the ADC unit on a stage with 5 degrees of freedom X,Y,Z, Z-tilt (in optical axis direction), and rotation. We aligned the ADC unit to the incoming collimated 633nm laser beam using back-reflected ghosts from the ADC1 different surfaces, see fig. \ref{fig:reflections}. We characterized the back reflections by the ray-tracing software and we were able to identify each spot we saw. In this way we used the rotation of these spot while rotating the motor to identify the rotational axis and measuring the tilt of the optics (in particular looking at the angular distance between the S3 reflected beam (Fig. \ref{fig:reflections}) with respect to the rotational axis. \\
Also, we inspected the spots' movement at a CCD positioned at the focal plane as the individual motors were rotated. The images stacked in a single frame of the spot with the motors at several different positions are visible in Fig. \ref{fig:rotationsatrecieving}.

\begin{itemize}
    \item We estimated a misalignment between the optical axis of the ADC1 quadruplet and the rotational axis of the ADC1 motor close to 1.5 degrees.
    \item We aligned the ADC1 quadruplet optical axis to the incoming beam axis through back reflections. Then we rotated the ADC1 motor 360 deg taking an image at the CCD placed at the focal plane (transmitted through ADC1 and ADC2) every 20deg, keeping ADC2 at fixed position. We see that the spot makes a circle at the CCD with a diameter of 30um (see Fig. 1 left-panel), this value was acceptable.
    \item In the same optical configuration as the previous point, we rotated the ADC2 motor 360 deg taking an image at the CCD (transmitted through ADC1 and ADC2) every 20deg, keeping ADC1 at fixed position. We see that the spot makes a circle at the CCD with a diameter of 150um (see Fig. 1 right-panel). Unfortunately, this value was too large to be accepted.
    \item Reversing the ADC position (that is ADC2 facing the incoming beam), we estimated a misalignment between the optical axis of the ADC2 quadruplet and the rotational axis of the ADC2 motor close to 1 degree. 
\end{itemize}

\section{METHODS FOR THE NEW ALIGNMENT}
\label{sec:METHODS}

\begin{figure}
    \centering
    \begin{subfigure}[b]{0.45\textwidth}
            \includegraphics[width=\textwidth]{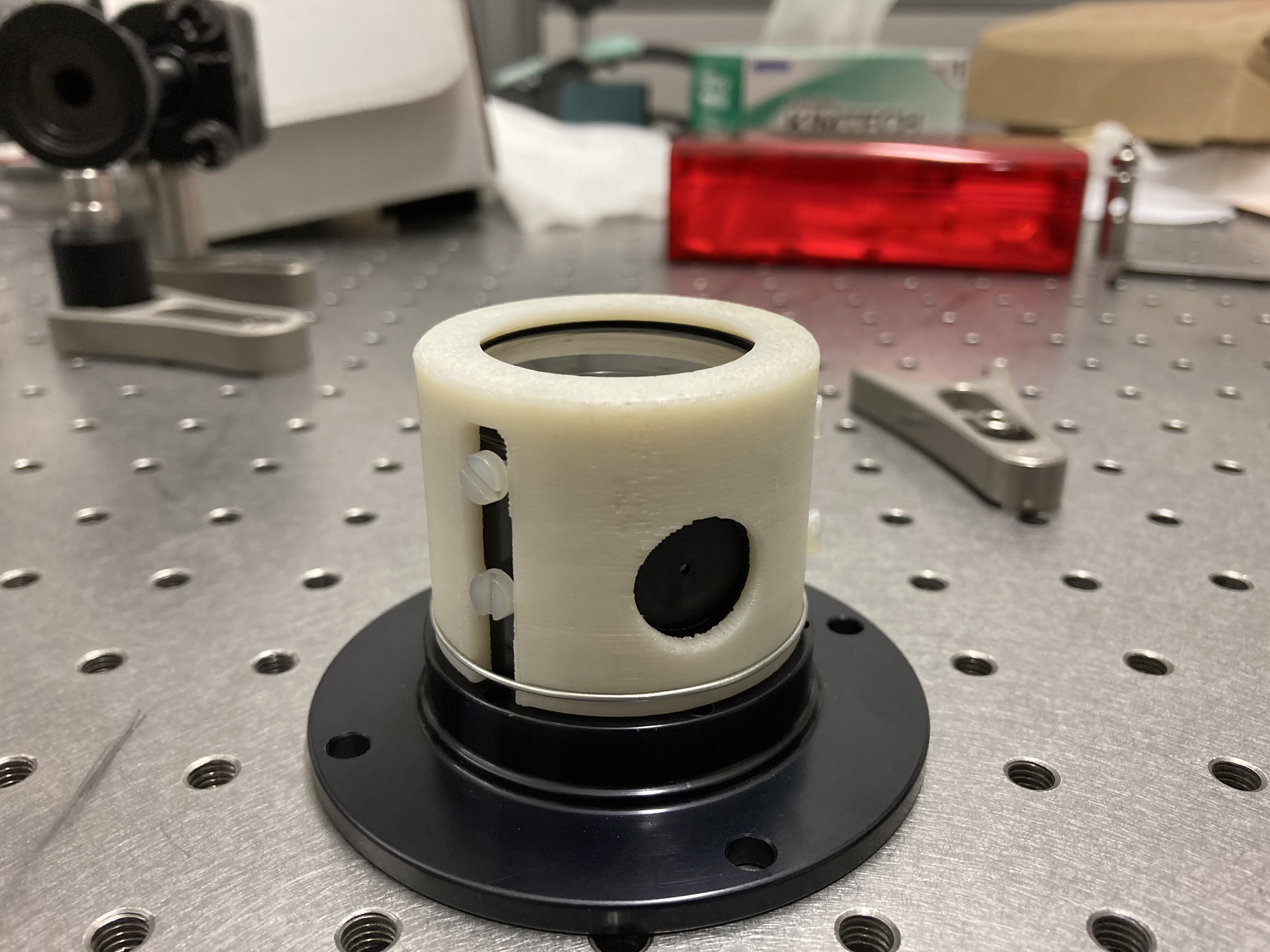}
    \caption{ADC1 with the 3D printed retainer}
    \label{fig:adc1_3dprint}
    \end{subfigure}
    \begin{subfigure}[b]{0.45\textwidth}
                \includegraphics[width=\textwidth]{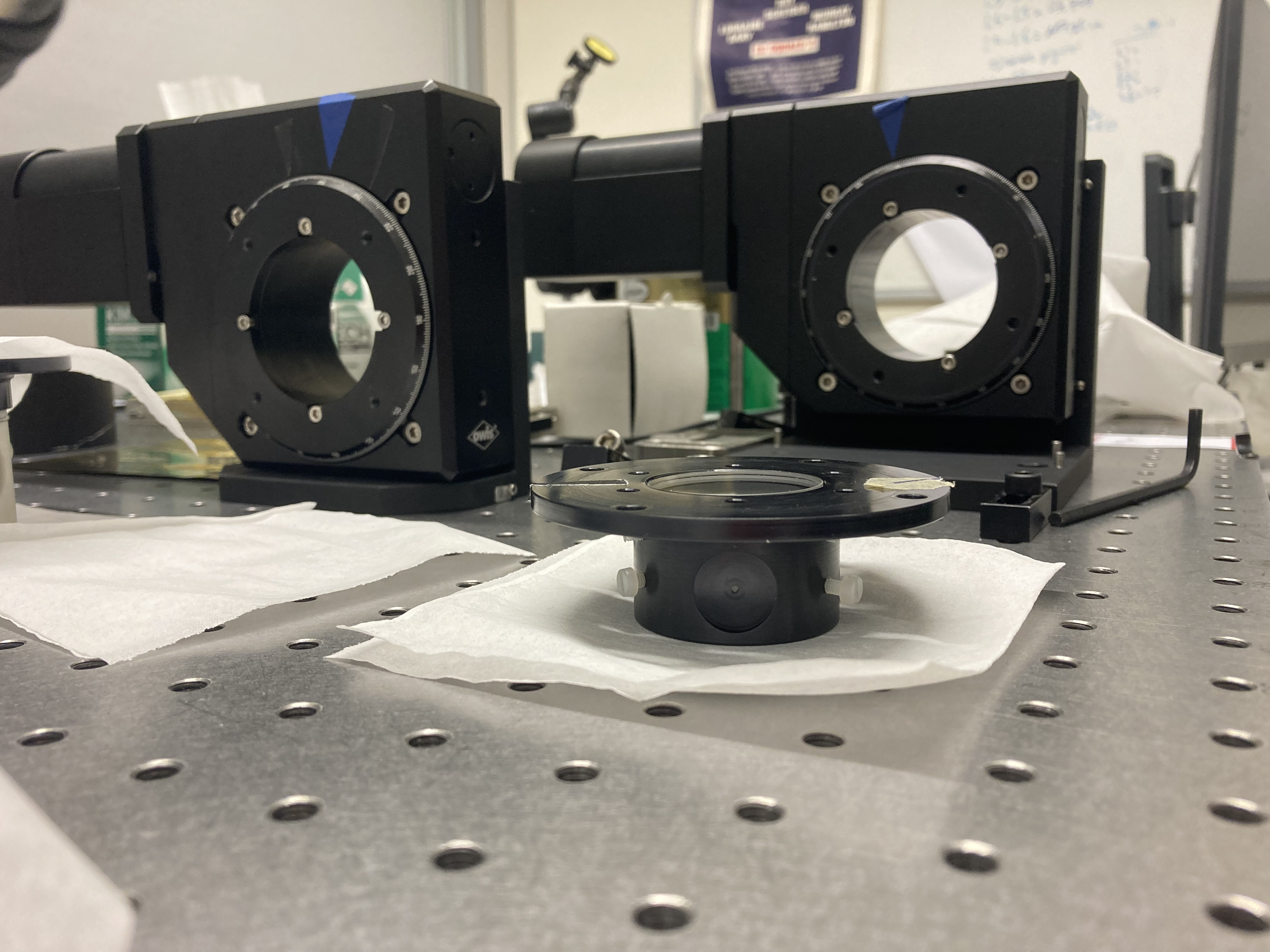}
    \caption{ADC2 with the teflon screws}
    \label{fig:adc2_teflonscrew}
    \end{subfigure}
    \caption{Barrels in the new alignment fashion}\label{fig:barrels}
\end{figure}

In order to control the tilt of the quadruplet inside the barrel we drilled threaded holes on two rows. The holes can host M3 teflon screws positioned 120° to each other, see Fig. \ref{fig:barrels}.  \\
We also produced two 3D printed holders to prevent any risk of falling off of the quadruplets, see Fig. \ref{fig:barrels} left panel. 

We proceeded with the single unit internal alignment. We used both back reflection from the various surfaces and transmission effect on a 633 nm laser beam and on a telescope simulator beam. We obtained acceptable results at the end of December 2021.\\
We built a set-up with both a laser reference and an on-axis telescope simulator feed by a Thorlabs SLS201L/M lamp with the possibility to insert narrow band filters. We aligned each subunit wrt the incoming beam using different autocollimation configurations.
\begin{figure}
    \centering
    \begin{subfigure}[t]{0.45\textwidth}
            \includegraphics[width=\textwidth]{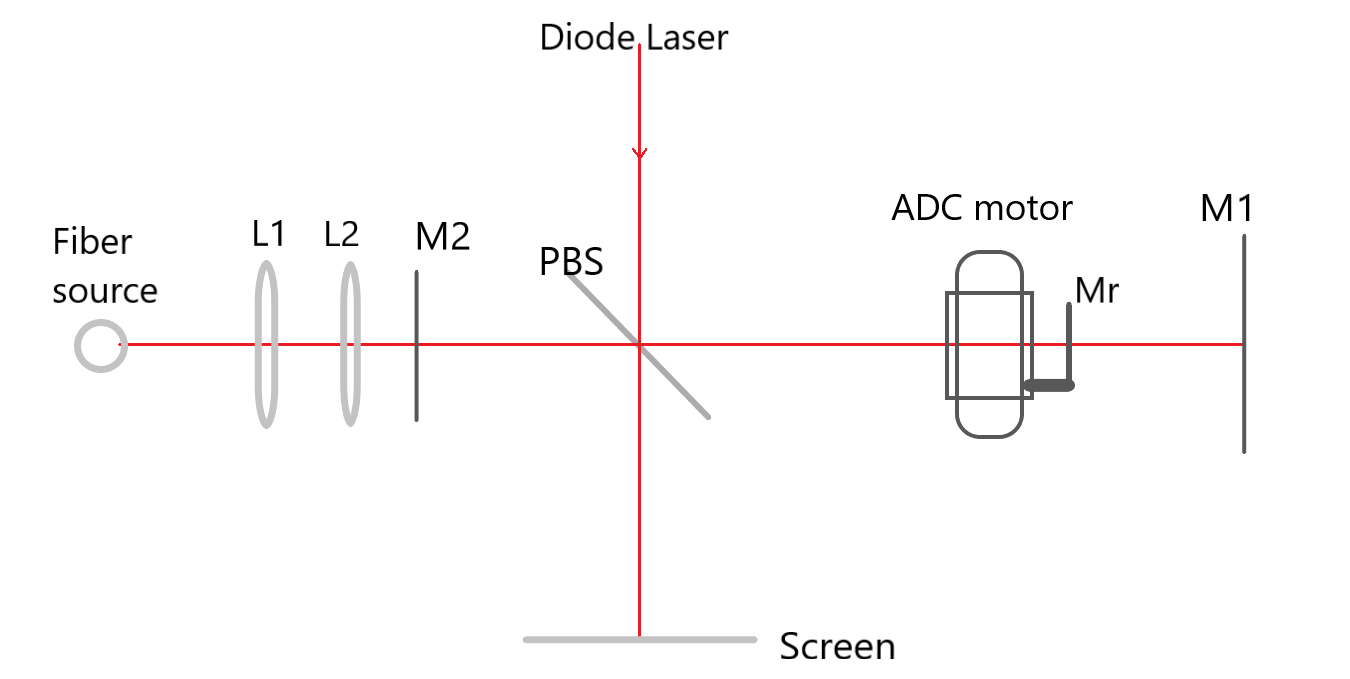}
    \caption{Taking care of the tilt of the motor axis}
    \label{fig:setupmirr}
    \end{subfigure}
    \begin{subfigure}[t]{0.45\textwidth}
                \includegraphics[width=\textwidth]{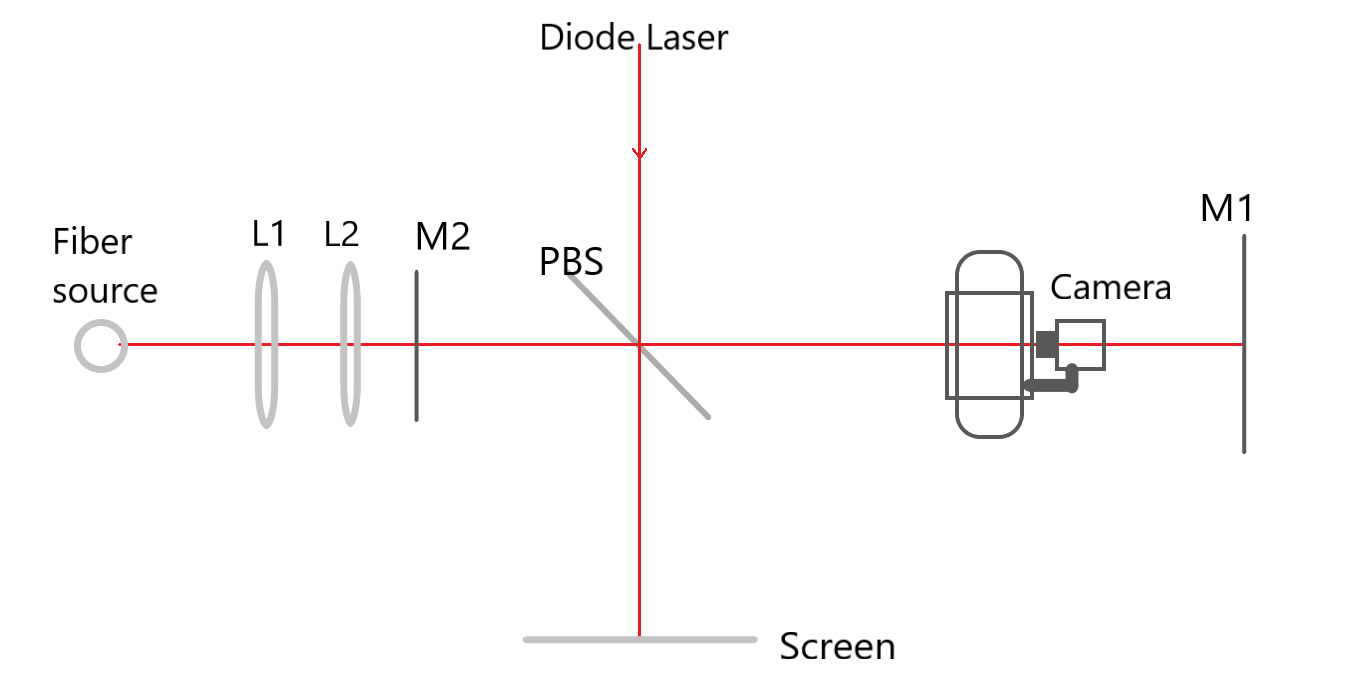}
    \caption{Taking care of the motor axis decenter}
    \label{fig:setupcamera}
    \end{subfigure}
    \begin{subfigure}[b]{0.8\textwidth}
                \includegraphics[width=\textwidth]{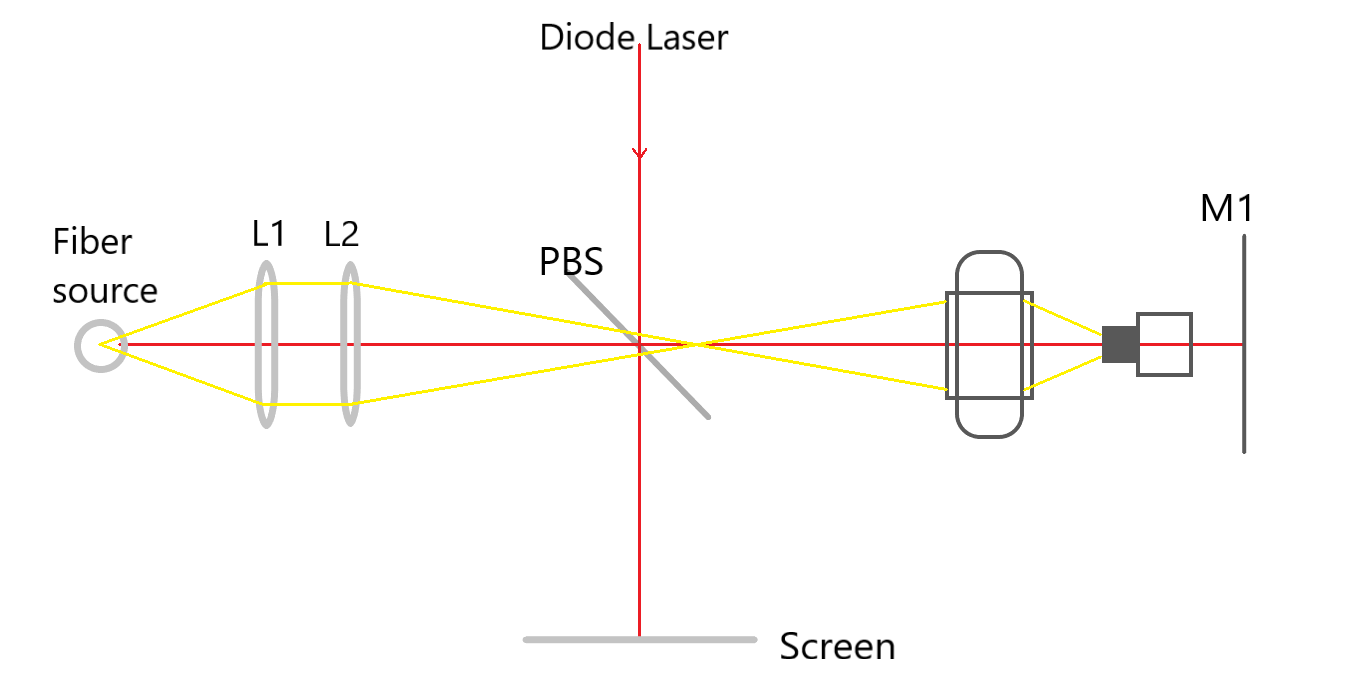}
    \caption{Aligning the telescope simulator}
    \end{subfigure}
    \caption{Re-Alignment Setup}\label{fig:reAlSetUp}
\end{figure}
A mirror and then a service camera were used screwed to the rotor to correct tilt and decenter of the motor, see Fig.  \ref{fig:benchsetup} for a picture of the bench set-up and \ref{fig:reAlSetUp} for a schematic representation of the setup. See figure to 
\begin{figure}
    \centering
    \includegraphics[width=0.5\textwidth]{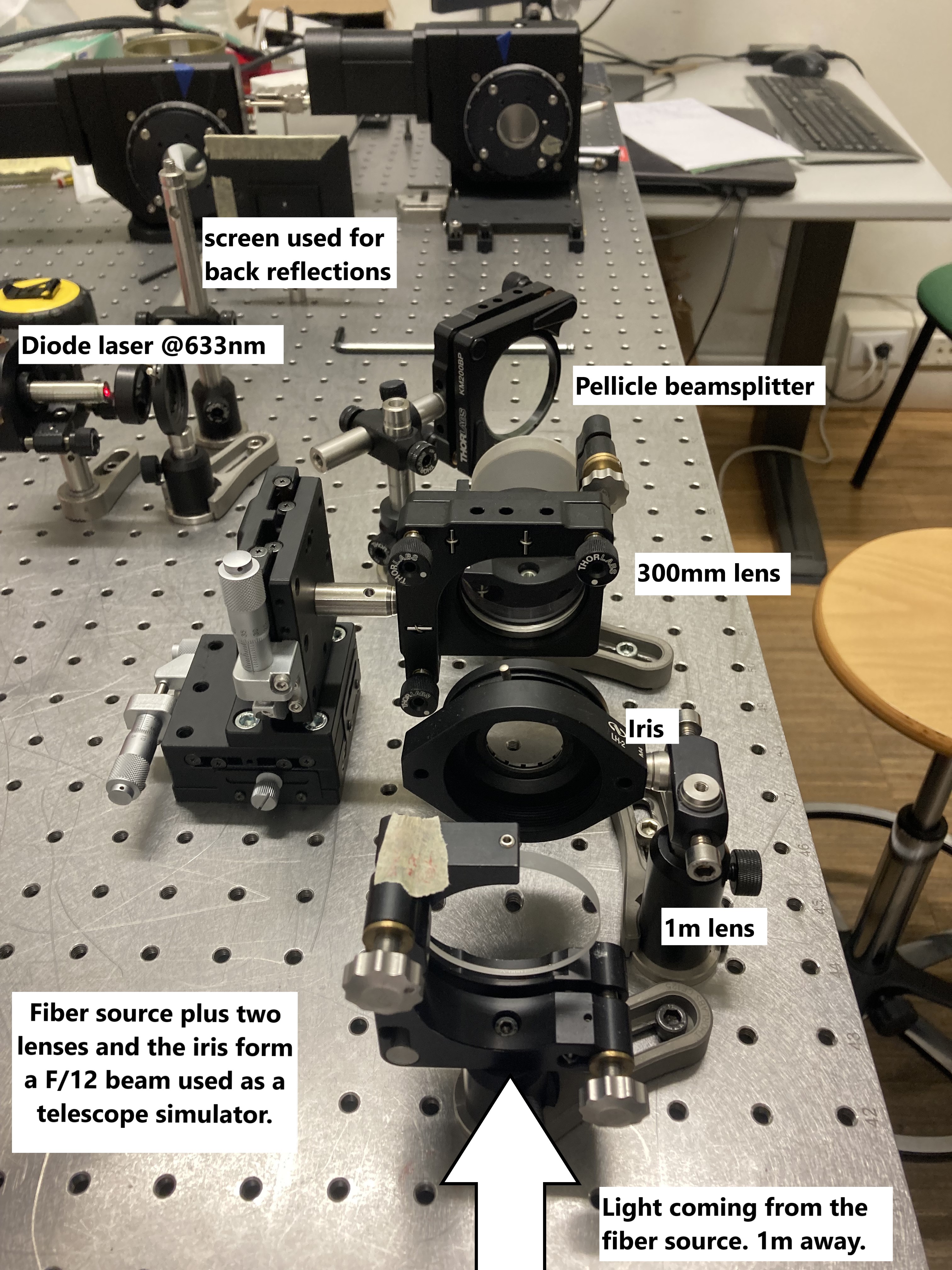}
    \caption{The optical bench set-up used for the realignment of the ADC}
    \label{fig:benchsetup}
\end{figure}
We then aligned the two ADC sub-units (quadruplets) in their modified barrels. In order to assess the alignment and optical performance of the sub-units and the whole ADC we used the following observables:
\begin{itemize}
    \item Laser back-reflection from the S3 planar surfaces (see Fig. \ref{fig:reflections}) of each quadruplets;
    \item Movement of the transmitted laser beam while rotating the motors, expected to be close to zero by design at 633nm;
    \item Dispersion introduced by the quadruplet measured by the diameter of the circle drawn by the on-focus spot using different narrow band filters;
    \item Psf quality, encircled energy and FWHM. 
\end{itemize}
\section{RESULTS AND CONCLUSIONS}
We performed a first evaluation of the optical quality in the temporary configuration where the quadruplets were hold in place by the strength of the six teflon screws. The results were good so we proceeded with the re-gluing process. We used an M3 2216 glue that took one week curing time at room temperature.\\
After a new centering of the barrels in the same way presented in the previous section we integrated the two subunits and and than we moved the ADCinside the Common Path of SOXS. ADC is the only powered optics in the UV-VIS arm inside the CP. \\
In table \ref{tab:alignment} we reported the data of alignment obtained for the two sub units.
\begin{table}[h]
\centering
\begin{tabular}{ll|l|l}
                           & Quadruplet \textbf{tilt} & Quadruplet \textbf{decenter} & Tolerances      \\ \hline \hline
\multicolumn{1}{l|}{ADC 1} & \textless{}50’’ & \textless{}20 µm    & 180’’ and 50 µm \\ \hline
\multicolumn{1}{l|}{ADC 2} & \textless{}50’’ & \textless{}20 µm    & 180’’ and 50 µm \\ \hline
\end{tabular}
\caption{Alignment of the single quadruplet with respect to the rotational axis of the ADC motors} \label{tab:alignment}
\end{table}
We measured the overall optical quality and the dispersion of the ADC. In Table \ref{tab:Optqulity} we summarized the data about FWHM and the encircled energy. PSFs have been acquired in white light for a grid of positions centred on on-axis.
For each position 10 images have been averaged; to the obtained image, the dark image (obtained averaging the 50 images of dark) has been subtracted.
Each obtained PSF has been fitted with a Gaussian. In the following table FWHM along one direction and the perpendicular one is reported as well as the e70,80 and 90 that are the radius containing the 70, 80 and 90\% of energy, in microns.
\begin{figure}[h]
    \centering
    \includegraphics[width=0.6\textwidth]{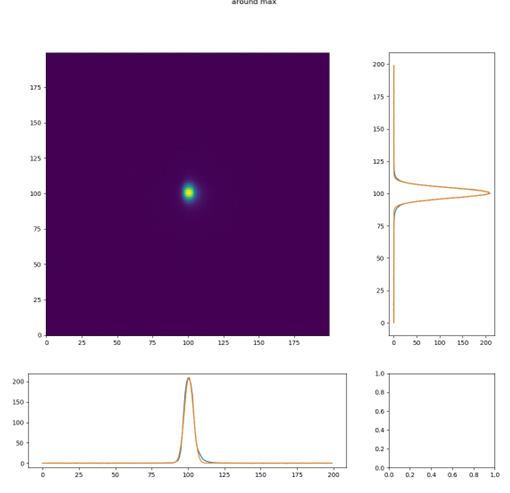}
    \caption{The PSF in white light at theoretically zero dispersion for the on-axis position. A residual elongation is visible, but in any case as visible in table \ref{tab:Optqulity} it is in very good agreement with the Zemax simulations. }
    \label{fig:my_label}
\end{figure}

\begin{table}[h]
\centering
\begin{tabular}{l|l|l}
\multicolumn{1}{c}{\textbf{}} & \multicolumn{1}{c}{\textbf{Zemax simulation}[$\mu m$]} & \multicolumn{1}{c}{\textbf{Mesured} [$\mu m$]} \\ \hline \hline
FWHM                          & 21                        & 12-16                       \\ \hline
E70                           & 10                        & 10-11                       \\ \hline
E80                           & 13                        & 11-14                       \\ \hline
E90                           & 17                        & 16-21                      \\ \hline
\end{tabular}
\caption{Optical quality measurements, comparison between the optical design nominal one and the measured} \label{tab:Optqulity}
\end{table}

\begin{figure}[h]
    \centering
    \begin{subfigure}[h]{0.45\textwidth}
            \includegraphics[width=\textwidth]{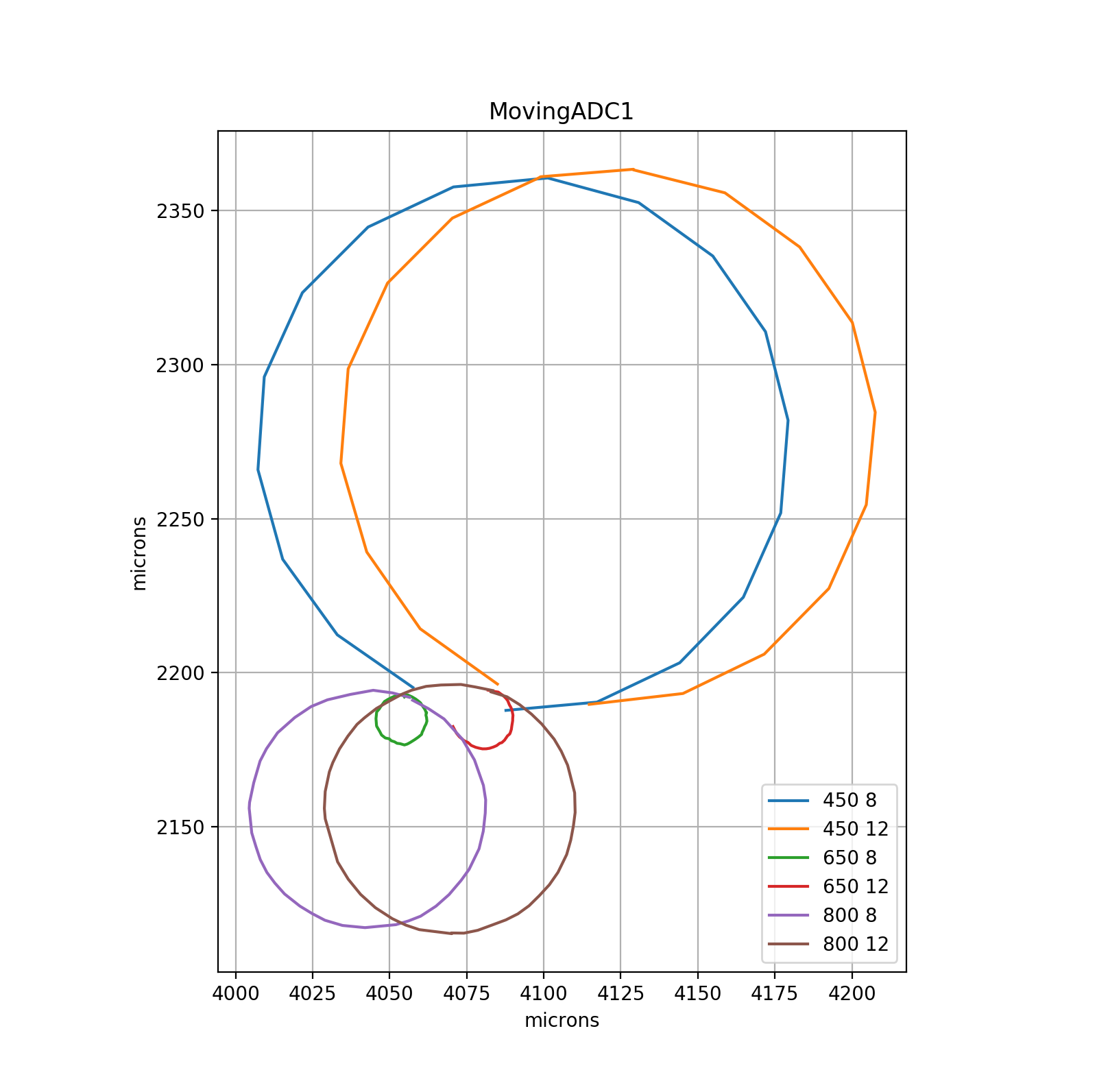}    
    \end{subfigure}
    \begin{subfigure}[h]{0.45\textwidth}
            \includegraphics[width=\textwidth]{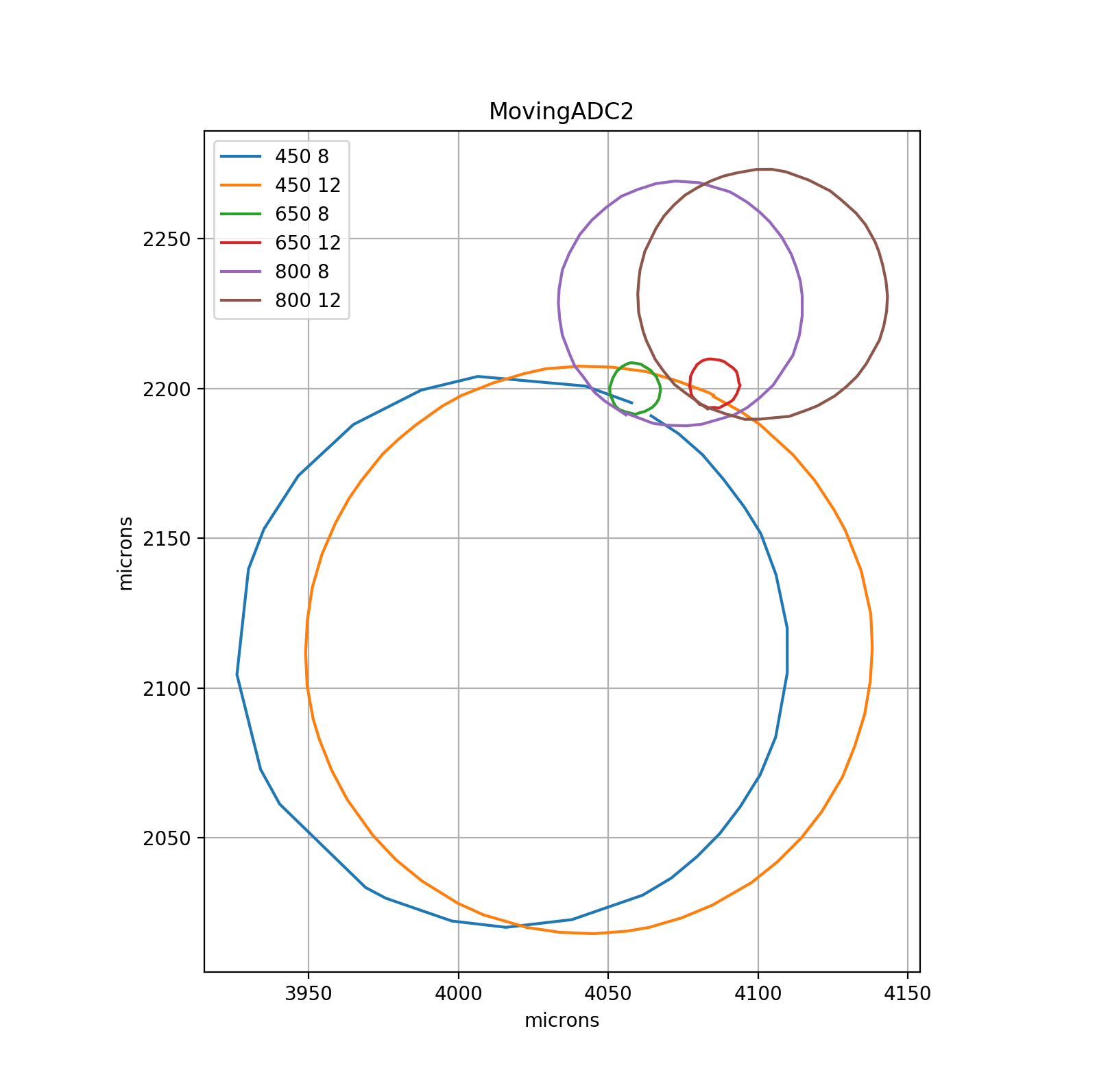}
    \end{subfigure}
    \caption{Dispersion at different wavelength}
    \label{fig:dispersion}
\end{figure}

In table \ref{tab:dispersion} are summarized the nominal and actual value for the dispersion at three different wavelength. We moved each motor at time and we took images of the monochromatic spot. The circle drawn by the spot as a radius that strictly depend on the dispersion angle (on sky working will be counter-dispersion) introduced by the ADC for the specific wavelength. In figure \ref{fig:dispersion}  the three circle for the 450nm, 650nm and 800nm are visible moving ADC1 or 2. An off-set in the diameter of the circle and so in the exit beam angle is present but very small and well manageable. 

\begin{table}[h]
\centering
\begin{tabular}{l|l|l|l}
Wavelength & ADC 1 {[}µm{]} & ADC 2 {[}µm{]} & Nominal {[}µm{]} \\ \hline \hline
450 nm     & 86.73          & 92.19          & 102              \\ \hline
650 nm     & 8.04           & 8.40           & 0                \\ \hline
800 nm     & 38.35          & 40.83          & 32               \\ \hline
\end{tabular}
\caption{Dispersion radius of the transmitted beam while rotating of of the motor at time}\label{tab:dispersion}
\end{table}

Summarizing, we can say that SOXS’ ADC is now well aligned inside the tolerances and we have defined a valid strategy for aligning a powered ADC.

\begin{itemize}
    \item Optical quality (encircled energy) and dispersion property are in accordance with ray tracing simulations;
    \item An alignment and verification strategy for a powered ADC is set;
    \item The new alignment inside the CP returns ~1\% error in F/\#.
\end{itemize}

\nocite{*}

\bibliography{main} 

\begin{thebibliography}{10}

\bibitem{Schipani16}
Schipani, P. et~al., ``The new {SOXS} instrument for the {ESO NTT},'' {\em
  Proc. SPIE} {\bf 9908},  990841 (2016).

\bibitem{Claudi18}
Claudi, R. et~al., ``The common path of {SOXS} ({S}on of {X}-{S}hooter),'' {\em
  Proc. SPIE} {\bf 10702},  107023T (2018).

\bibitem{Sanchez18}
Sanchez, R.~Z. et~al., ``Optical design of the {SOXS} spectrograph for {ESO
  NTT},'' {\em Proc. SPIE} {\bf 10702},  1070227 (2018).

\bibitem{Biondi20}
Biondi, F. et~al., ``The {AIV} strategy of the common path of {S}on of
  {X-S}hooter,'' {\em Proc. SPIE} {\bf 11447},  114476P (2020).

\bibitem{Rubin18}
Rubin, A. et~al., ``{MITS}: the {M}ulti-{I}maging {T}ransient {S}pectrograph
  for {SOXS},'' {\em Proc. SPIE} {\bf 10702},  107022Z (2018).

\bibitem{Schipani18}
Schipani, P. et~al., ``{SOXS}: a wide band spectrograph to follow up
  transients,'' {\em Proc. SPIE} {\bf 10702},  107020F (2018).

\bibitem{Vitali18}
Vitali, F. et~al., ``The {NIR} spectrograph for the new {SOXS} instrument at
  the {NTT},'' {\em Proc. SPIE} {\bf 10702},  1070228 (2018).

\bibitem{Brucalassi18}
Brucalassi, A. et~al., ``The acquisition camera system for {SOXS} at {NTT},''
  {\em Proc. SPIE} {\bf 10702},  107022M (2018).

\bibitem{Cosentino18}
Cosentino, R. et~al., ``The vis detector system of {SOXS},'' {\em Proc. SPIE}
  {\bf 10702},  107022J (2018).

\bibitem{Ricci18}
Ricci, D. et~al., ``Architecture of the {SOXS} instrument control software,''
  {\em Proc. SPIE} {\bf 10707},  107071G (2018).

\bibitem{Capasso18}
Capasso, G. et~al., ``{SOXS} control electronics design,'' {\em Proc. SPIE}
  {\bf 10707},  107072H (2018).

\bibitem{Aliverti18}
Aliverti, M. et~al., ``The mechanical design of {SOXS} for the {NTT},'' {\em
  Proc. SPIE} {\bf 10702},  1070231 (2018).

\bibitem{Biondi18}
Biondi, F. et~al., ``The assembly integration and test activities for the new
  {SOXS} instrument at {NTT},'' {\em Proc. SPIE} {\bf 10702},  107023D (2018).

\bibitem{Young20}
Young, D. et~al., ``The {SOXS} data reduction pipeline,'' {\em Proc. SPIE} {\bf
  11452},  114522D (2020).

\bibitem{Kuncarayakti20}
Kuncarayakti, H. et~al., ``Design and development of the {SOXS} calibration
  unit,'' {\em Proc. SPIE} {\bf 11447},  1144766 (2020).

\bibitem{Schipani20}
Schipani, P. et~al., ``Development status of the {SOXS} spectrograph for the
  {ESO-NTT} telescope,'' {\em Proc. SPIE} {\bf 11447},  1144709 (2020).

\bibitem{Genoni20}
Genoni, M. et~al., ``{SOXS} {E}nd-to-{E}nd simulator: development and
  applications for pipeline design,'' {\em Proc. SPIE} {\bf 11450},  114501B
  (2020).

\bibitem{Colapietro20}
Colapietro, M. et~al., ``Progress and tests on the {I}nstrument {C}ontrol
  {E}lectronics for {SOXS},'' {\em Proc. SPIE} {\bf 11452},  1145225 (2020).

\bibitem{Vitali20}
Vitali, F. et~al., ``The development status of the {NIR} arm of the new {SOXS}
  instrument at the {ESO/NTT} telescope,'' {\em Proc. SPIE} {\bf 11447},
  114475N (2020).

\bibitem{Ricci20}
Ricci, D. et~al., ``Development status of the {SOXS} instrument control
  software,'' {\em Proc. SPIE} {\bf 11452},  114522Q (2020).

\bibitem{Rubin20}
Rubin, A. et~al., ``Progress on the {UV-VIS} arm of {SOXS},'' {\em Proc. SPIE}
  {\bf 11447},  114475L (2020).

\bibitem{Cosentino20}
Cosentino, R. et~al., ``Development status of the {UV-VIS} detector system of
  {SOXS} for the {ESO-NTT} telescope,'' {\em Proc. SPIE} {\bf 11447},  114476C
  (2020).

\bibitem{Brucalassi20}
Brucalassi, A. et~al., ``Final design and development status of the acquisition
  and guiding system for {SOXS},'' {\em Proc. SPIE} {\bf 11447},  114475V
  (2020).

\bibitem{Sanchez20}
Sanchez, R.~Z. et~al., ``{SOXS}: effects on optical performances due to gravity
  flexures, temperature variations, and subsystems alignment,'' {\em Proc.
  SPIE} {\bf 11447},  114475F (2020).

\bibitem{Claudi20}
Claudi, R. et~al., ``Operational modes and efficiency of {SOXS},'' {\em Proc.
  SPIE} {\bf 11447},  114477C (2020).

\bibitem{Aliverti20}
Aliverti, M. et~al., ``Manufacturing, integration and mechanical verification
  of {SOXS},'' {\em Proc. SPIE} {\bf 11447},  114476O (2020).

\end{thebibliography}
\bibliographystyle{spiebib} 

\end{document}